\documentclass[prd,showpacs,preprintnumbers,amsmath,amssymb]{revtex4}

\usepackage{graphicx}
\usepackage{dcolumn}
\usepackage{bm}

\def\be{\begin{eqnarray}}
\def\ee{\end{eqnarray}}
\def\bea{\begin{eqnarray}}
\def\eea{\end{eqnarray}}
\def\kT{{\bf k}_\perp}

\def\sT{{\bf s}_\perp}
\def\yT{{\bf y}_\perp}
\def\xT{{\bf x}_\perp}

\def\0T{{\bf 0}_\perp}
\begin{document}


\title{Sivers Mechanism for Gluons}

\author{Matthias Burkardt}
 \affiliation{Department of Physics, New Mexico State University,
Las Cruces, NM 88003-0001, U.S.A.}

\date{\today}

\begin{abstract}
We study T-odd gauge invariantly defined T-odd unintegrated gluon 
densities in QCD. 
The average $\perp$ momentum of the gluons in a $\perp$
polarized nucleon can be nonzero due to the Sivers mechanism.
However, we find that the sum of the transverse momenta 
due to the Sivers mechanism
from gluons and all quark flavors combined vanishes, i.e.
$\langle {\bf k_{\perp,g}}\rangle + \sum_q
\langle {\bf k_{\perp,q}}\rangle =0$.
\end{abstract}

\maketitle
\section{Introduction}
Recent experiments performed by the HERMES collaboration have 
demonstrated that a small but nonzero Sivers mechanism for 
transverse single-spin asymmetries exists \cite{hermes}.
Although the presence of such effects has been conjectured 
long ago \cite{sivers}, only more recent studies
\cite{hwang} have clarified the role of final and initial state 
interactions as a source of such T-odd asymmetries at the parton 
level and have thus demonstrated that a nonzero Sivers mechanism is 
possible in QCD. Following these insights, several model estimates
for the Sivers asymmetry have been made \cite{model} but only 
few theoretical 
constraints exist regarding the anticipated magnitude or even the
sign of the resulting asymmetries in QCD. The main purpose of this 
letter is to derive a constraint on the Sivers mechanism for gluons.

For the final (initial) state interactions, previous results for 
the Sivers mechanism for quarks can
be summarized in a relatively compact form by means 
parton densities with $T-odd$ contributions. 
In the case of unintegrated quark densities
one finds \cite{collins,ji,boer}
\be
q(x,\kT,\sT) &=& \! \int \frac{dy^- d^2\yT }{16\pi^3}
e^{-ixp^+y^-+i\kT \cdot \yT } \label{eq:P}
\left\langle p \left|\bar{q}(y^-,\yT) \gamma^+
\left[y^-,\yT;\infty^-,\yT \right]
\left[\infty^-,\0T;0^-,\0T\right] q(0)\right|p\right\rangle .
\ee
We use light-front (LF) coordinates, which are defined as:
$y^\mu =(y^+,y^-,\yT)$, with $y^\pm = (y^0\pm y^3)/\sqrt{2}$.
In all correlation functions, $y^+=0$ and we therefore do not
explicitly show the $y^+$ dependence.
The path ordered Wilson-line operator from the point $y$ to infinity
is defined as
\bea
\left[\infty^-,\yT;y^-,\yT \right] =
P\exp \left(-ig\int_{y^-}^\infty dz^- A^+(y^+,z^-,\yT)
\right).
\label{W1}
\eea
The specific choice of path in Ref. \cite{collins} reflects the
FSI (ISI) of the active quark in an eikonal approximation.
The complex phase in Eq. ({\ref{eq:P}) is reversed under 
time-reversal and therefore T-odd PDFs may exist \cite{collins}.
It should be emphasized that the presence of small $x$ and UV
divergences and the necessary renormalizations
prevent us from literally interpreting these densities as
number densities \cite{collins2}. However, in this work we will
only study the resulting average value of $\kT$ integrated over all 
$x$, where most of these divergences are expected to cancel.

Naively, the single spin asymmetry seems to be absent in light-cone 
gauge $A^+=0$, since the Wilson lines in Eq. (\ref{eq:P}) are
in the $x^-$ direction and therefore $\int dz^- A^+ =0$.
Without the phase factor any single spin asymmetry vanishes due to
time reversal invariance.
This apparent puzzle has been resolved in Ref. \cite{ji}, where it 
has been emphasized that a truly gauge invariant definition for
unintegrated parton densities requires closing the gauge link at
$x^-=\infty$, i.e. a fully gauge invariant version of Eq. 
(\ref{eq:P}) reads (Fig. \ref{fig:staple})
\be
q(x,\kT,\sT) &=& \!\!\int \frac{dy^- d^2\yT }{16\pi^3}
e^{-ixp^+y^-+i\kT \cdot \yT } 
\left\langle p \left|\bar{q}(y^-,\yT) \gamma^+
\left[y^-,\yT;\infty^-,\yT \right] 
\left[\infty^-,\yT,\infty^-,\0T\right]
\left[\infty^-,\0T;0^-,\0T\right] q(0)\right|p\right\rangle .
\nonumber\\
\label{eq:P2}
\ee
\begin{figure}
\unitlength1.cm
\begin{picture}(10,5)(5.3,21.5)
\includegraphics{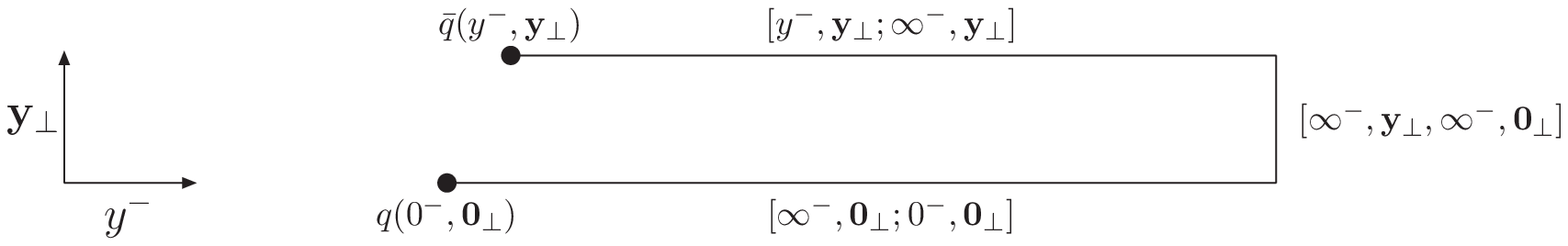}
\end{picture}
\caption{Illustration of the gauge links in Eq. (\ref{eq:P2}).}
\label{fig:staple}
\end{figure}  
In all commonly used gauges, except the light-cone 
gauge, the gauge link at $x^-=\infty$ is not expected to
contribute to the matrix element, since the gauge fields are 
expected to fall off rapidly enough at $\infty$. However, this is not
true in light-cone gauge and therefore it has been suggested in 
Ref. \cite{ji} that, in light-cone gauge, the entire single-spin 
asymmetry arises from the phase due to the gauge link at 
$x^-=\infty$
\be
q(x,\kT,\sT) &=& \int \frac{dy^- d^2\yT }{16\pi^3}
e^{-ixp^+y^-+i\kT \cdot \yT } \label{eq:P3}
\left\langle p \left|\bar{q}(y^-,\yT) \left[\infty^-,\yT,\infty^-,\0T\right]
\gamma^+ q(0)\right|p\right\rangle 
\quad \quad \quad \quad (A^+=0).
\ee
The main purpose of this letter is to generalize these results to
T-odd parton distributions for the glue and to discuss the
consequences for the net Sivers effect.

\section{The Unintegrated Gluon Density}
The naive (not gauge invariant and not accounting for T-odd effects)
definition for the unintegrated unpolarized (i.e. gluons unpolarized)
gluon density reads (see also Refs. \cite{rodrigues,vogel}, where a
decomposition into Lorentz components is provided)
\bea
x g(x,\kT,\sT) &\equiv&  \label{mulders}
\int \frac{dy^- d^2\yT}{8\pi^3p^+}
e^{-ixp^+y^-+i\kT\cdot \yT} \left\langle p\left|
tr\left( F^{+i}(y)F^{+i}(0)\right)\right|p\right\rangle
,\nonumber
\eea
where $F^{\mu \nu }$ is the gluon field strength tensor.
Throughout this paper, we consider states $|p\rangle$, which are
polarized in the $\perp$ direction, but suppress the spin label in 
the state in order to shorten the notation.

In analogy with gauge invariant parton densities for the quarks,
we  now introduce gauge invariant gluon densities by augmenting
Eq. (\ref{mulders}) with appropriate (future-pointing) Wilson lines
\bea
x g(x,\kT,\sT)  \label{mulders2}
&=& \int \frac{dy^- d^2\yT}{8\pi^3p^+}
e^{-ixp^+y^-+i\kT\cdot \yT}
\left\langle p\left|tr\left(
\hat{F}^{+i}(y)
\left[\infty^-,\yT;\infty^-,\0T \right]
\hat{F}^{+i}(0)
\left[\infty^-,\0T;\infty^-,\yT \right]
\right)\right|p\right\rangle
, 
\eea
where
\bea
\hat{F}^{+i}(y)&=&
\left[\infty^-,\yT;y^-,\yT\right] F^{+i}(y)
\left[y^-,\yT;\infty^-,\yT \right] .
\eea
The particular choice of Wilson line (here future-pointing)
depends on the process to which the gluon distributions are
applied. Throughout this paper we focus on distributions defined
with future-pointing lines. Past-pointing lines can be trivially
obtained by a time-reversal operation and yield the same result,
but with opposite signs for the transverse single-spin asymmetry.

Similarly to the quark case, the gauge link at 
$x^-=\pm \infty$ is expected to contribute only in the
light-cone gauge, where the gauge links along the light-cone
vanish and we can thus replace $\hat{F}^{+i}
\rightarrow F^{+i}$ in Eq. (\ref{mulders2}). In the following
we will work in light-cone gauge, where
$
F^{+i} = \partial_- A^i
$
and therefore
\bea 
g(x,\kT,\sT) =
i\int \frac{dy^- d^2\yT}{8\pi^3} e^{-ixp^+y^-+i\kT\cdot \yT}
 \left\langle p\left|tr\left(
A_i(y)
\left[\infty^-,\yT;\infty^-,\0T \right]
\partial_- A_i(0)
\left[\infty^-,\0T;\infty^-,\yT \right]
\right)\right|p\right\rangle
\label{g2}
\eea

\section{Average Transverse Momentum}
In order to evaluate the average transverse momentum we multiply
Eq. (\ref{g2}) by $\kT$ and integrate over $d^2\kT$. The factor
$\kT$ is replaced by a derivative w.r.t. $\yT$. The term
where this $\perp$ derivative acts on $A_i(y)$ does not
contribute in the end because of time reversal invariance.
The only relevant contribution arises when the derivative acts on
the gauge link at $y^-=\infty$, yielding
\bea
\bar{\bf k}_{\perp g}(x) &\equiv& 
\int d^2\kT \kT g(x,\kT,\sT)\label{kbar}
=g f^{abc}\int \frac{d y^-}{4\pi} e^{-ixp^+y-}
\left\langle p\left| 
A_{i,b}(y^-)\partial_-A_{i,c}(0)A_{\perp,a}
(\infty^-,\0T)\right|p\right\rangle .
\eea
Time reversal transformations reverse the orientation of the
gauge string and we can therefore replace $A_{\perp,a}(\infty^-,\0T)
\rightarrow {\bf \alpha}_{\perp,a}(\0T)$, where
\be
{\bf \alpha}_{\perp a}(\0T)\equiv \frac{1}{2}
\left(
{\bf A_{\perp a}}(\infty^-,\0T) -{\bf A_{\perp a}}(-\infty^-,\0T)
\right).
\ee
In particular we find for the net transverse momentum carried by the
gluons due to the Sivers effect
\be
\langle {\bf k}_{\perp g}\rangle
&\equiv& \int dx \int d^2\kT g(x,\kT,\sT) =
\frac{g}{2p^+} 
\left\langle p \left|f^{abc}
\left(A_{i,b}(y^-),\partial_-A_{i,c}(0)\right]{\bf \alpha}_{\perp,a}
\right|p\right\rangle .
\label{gSivers}
\ee
Eq. (\ref{gSivers}) is invariant under residual gauge 
transformations, which keep $A^+=0$.
One of the applications of this result lies in thee fact that 
Eq. (\ref{kbar}) can be more directly evaluated in
models (or calculations) of the light-cone wave function of hadrons
than the starting equation (\ref{mulders2}). However, the main
significance of this result will become more evident below when we 
combine Eq. (\ref{kbar}) with the analog relation for quarks.

In Ref. \cite{me} it was found that, in light-cone gauge, 
the average transverse momentum (integrated over $x$)
of the active quark resulting from this definition can be 
related to the correlation between the colored quark density
and the transverse gauge field at $x^-=\pm \infty$
\bea
\left\langle {\bf k}_{\perp q}\right\rangle &=&
\int dx \int d^2\kT q(x,\kT,\sT) \kT \label{kAA}
=
-\frac{g}{2p^+} 
\left\langle p \left|\bar{q}(0)  \gamma^+\frac{\lambda_a}{2}q(0)
{\bf \alpha}_{\perp a}(\0T) \right|p\right\rangle .
\eea
From the condition that the light-cone energy is finite at 
$x^-=\pm \infty$, one can derive several constraints on the gauge
field at $x^-=\pm \infty$ \cite{me}. First it must be pure gauge
$F_{12}^a\left|p\right\rangle=0$, i.e. with appropriate boundary
conditions (e.g. anti-symmetric) this implies
\be
\left( \partial_1 \alpha_{2,a} -\partial_2 \alpha_{1,a} +gf^{abc}
\alpha_{1,b}\alpha_{2,c} \right) \left|p\right\rangle=0.
\ee
Secondly, the condition that $F_{+-}(\pm \infty^-,\xT)|p\rangle=0$ 
yields \cite{me,raju}
\bea
\partial_i \alpha_{i,a}(\xT) \left|p\right\rangle
&=& -\rho_a(\xT)\left|p\right\rangle  
\label{nonabel1}
\eea
where $\rho_a(\xT)=\int dx^- J_a(x^-,\xT)$ and 
\bea
J_a(x^-,\xT) &=& -gf_{abc} A^i_b \partial_-A^i_c
+ g \sum_q \bar{q} \gamma^+\frac{\lambda_a}{2}q .
\label{nonabel2}
\eea
If we now add up the total Sivers effect for quarks and gluons,
we thus find
\bea
\label{grandtotal}
\left\langle {\bf k}_{1,g}\right\rangle +\sum_q
\left\langle {\bf k}_{1,q}\right\rangle
=
-\frac{g}{2p^+} 
\left\langle p \left|J_a(0) {\bf \alpha}_{\perp a}(\0T)
\right|p\right\rangle ,
\eea
This result is very interesting because it contains the same color 
charge density $J_a$ (\ref{nonabel2}) that also appears in the 
constraint equation for
${\bf \alpha}_{\perp a}$ (\ref{nonabel1}). In the following we will
demonstrate that the total transverse momentum (\ref{grandtotal})
vanishes identically. For this purpose we first use translation
invariance in the $y^-$ direction to replace $J_a(0)$ by
$\frac{1}{L} \rho_a(\0T) = \frac{1}{L} \partial_i \alpha_{i,a}(\0T)$ 
in Eq. (\ref{grandtotal}), where $L$ is the
length in $x^-$ of a fictitious `box' that we introduce here for
regularization purposes. For a target that is polarized in the 
$\hat{y}$ direction we need to consider only the momentum in
the $\hat{x}$ direction since the momentum in the $\hat{y}$ direction
vanishes already for each piece separately.
Combining the above results for the $\hat{x}$ component we thus find
\bea
\label{zero1}
\left\langle {\bf k}_{1,g}\right\rangle +\sum_q
\left\langle {\bf k}_{1,q}\right\rangle &=& -\frac{g}{2p^+L} 
\left\langle p \left| \left(\partial_i \alpha_{i,a}\right)
\alpha_{1,a} \right|p\right\rangle
= -\frac{g}{2p^+L} 
\left\langle p \left| \left(\partial_2 \alpha_{2,a}\right) 
\alpha_{1,a} \right|p\right\rangle
=\frac{g}{2p^+L} 
\left\langle p \left|  \alpha_{2,a}\partial_2 \alpha_{1,a} 
\right|p\right\rangle
\eea
where we first used that the forward matrix element of
$\left(\partial_1 \alpha_{1,a}\right) \alpha_{1,a} = \frac{1}{2}\partial_1
\left(\alpha_{1,a}\alpha_{1,a}\right)$ vanishes and then, for the same reason,
we replaced $\left(\partial_2 \alpha_{2,a}\right) \alpha_{1,a}$ by
$-\alpha_{2,a}\partial_2 \alpha_{1,a}$. We now use the fact that
${\bf \alpha}_\perp$ is pure gauge, i.e. acting on finite energy 
states we must have
\bea
\partial_2\alpha_{1,a} \left|p\right\rangle
= \left[ \partial_1\alpha_{2,a} 
-g f^{abc}\alpha_{1,b} \alpha_{2,c}\right]
\left|p\right\rangle.
\eea
Upon inserting this result in Eq. (\ref{zero1}) and utilizing
again the fact that the forward matrix element of a total derivative
$\alpha_{2,a} \partial_1\alpha_{2,a} = \frac{1}{2}\partial_1
\left(\alpha_{2,a}\alpha_{2,a}\right)$ vanishes we finally find
\bea
\label{zero2}
\left\langle {\bf k}_{1,g}\right\rangle +\sum_q
\left\langle {\bf k}_{1,q}\right\rangle
 &=& -\frac{g^2}{p^+L} 
\left\langle p \left| \alpha_{1,a} 
f^{abc}\alpha_{1,b} \alpha_{2,c} \right|p\right\rangle =0
\eea
due to the antisymmetry of the $SU(N)$ structure constant.
While the presence of the final state interactions allows a nonzero
Sivers mechanism for the gluons as well as for each quark flavor,
the net
transverse momentum of all partons (quarks plus gluons) resulting
from the Sivers effect must vanish.
In Ref. \cite{me} a similar result was derived for QED. 
In the QED case the physics of this result is much more transparent
because there one can solve the ``finiteness conditions'' 
explicitly and we refer the reader to that paper for more details.

The Sivers effect arises because a parton can be transversely 
deflected when it traverses the Lorentz contracted gauge field from 
the spectators. The physical interpretation of our above result is
that the transverse forces which the partons can exert on each other
due tue the gauge field interactions cancel each other when we sum
over all partons. Such a result is very familiar from Newton's 
$3^{rd}$
law where action=reaction implies that the net force on a 
multi-particle system with only internal forces vanishes. What we
observe here is a similar effect, but we will not try to exploit
this classical analogy here any further.

\section{Summary}
We have introduced gauge invariant unintegrated gluon densities
with future-pointing gauge strings and derived relations between
the resulting gluon-Sivers effect and a correlator between the
gluons density and the transverse gauge field at $x^-=\pm \infty$.
While we considered here only unintegrated gluon densities
with future-pointing gauge strings, the results can be easily
generalized to distributions with past-pointing gauge strings
by a time-reversal transformation. In particular the spin asymmetry 
simply changes sign for past-pointing strings. Whether future- or
past-pointing strings should be used depends is not arbitrary  
but depends on the experimental conditions in which they are used.
The results from this paper, 
with appropriate signs, should be applicable to either case.

While the net transverse momentum of the gluons arising from the
Sivers effect can be nonzero, we demonstrated that the net 
transverse Sivers momentum from all quark flavors plus the 
gluonic piece combined vanishes.
Such a result may be intuitively not so surprising, but 
it is nevertheless nontrivial when one starts from gauge invariantly
defined unintegrated parton densities, where transverse momentum
conservation is not evident. In fact, first ans\"atze to parameterize
the gluon-Sivers distribution \cite{vogel} did not utilize such a
constraint.

The main application of our result is that it provides an
additional constraint on parameterizations of Sivers distributions
allows to relate the
average Sivers effect for different experiments (which may probe
different flavors or the glue) to one another.

{\bf Acknowledgements:}
I would like to thank J.Collins and W.Vogelsang for stimulating 
discussions. This work was supported by the DOE under grant number 
DE-FG03-95ER40965.

\bibliography{gSivers.bbl}

\begin{thebibliography}{3}
\bibitem{hermes} N.C.R. Makins (HERMES collaboration), talk at eRHIC
workshop, BNL, Jan. 2004; see also http://www-hermes.desy.de
\bibitem{sivers} D.W. Sivers, Phys.\ Rev.\ D {\bf 43}, 261 (1991).
\bibitem{hwang} S.J. Brodksy, D.S. Hwang, and I. Schmidt,
Phys.\ Lett.\ B {\bf 530}, 99 (2002).
\bibitem{model} M. Burkardt, hep-ph/0302144;
F. Yuan, Phys. Lett. {\bf B575}, 45 (2003);
A. Bachetta, A. Sch\"afer, and J.-J. Yang, Phys.
Lett. B {\bf 578}, 109 (2004).
\bibitem{ji} X. Ji and F. Yuan, Phys. Lett. B {\bf 543}, 66 (2002);
A. Belitsky, X. Ji, and F. Yuan, Nucl. Phys. B {\bf 656}, 165 (2003).
\bibitem{collins} J.C. Collins, Phys. Lett. B {\bf 536}, 43 (2002);
D. Boer, P.J. Mulders, and F. Pijlman, Nucl. Phys. B {\bf 667},
201 (2003); see also R.D. Tangerman and P.J. Mulders, 
Phys. Rev. D {\bf 51}, 3357 (1995).
\bibitem{boer} D. Boer, P.J. Mulders, and F. Pijlman, Nucl. Phys.
B {\bf 667}, 201 (2003). 
\bibitem{collins2} J.C. Collins, Acta Phys.\ Polon.\ B {\bf 34},
3103 (2003).
\bibitem{vogel} D. Boer and W. Vogelsang, hep-ph/0312320.
\bibitem{rodrigues} P.J. Mulders and J. Rodrigues, Phys.\ Rev.\ D
{\bf 63}, 094021 (2001). 
\bibitem{me} M. Burkardt, hep-ph/0311013.
\bibitem{raju} L. McLerran and R. Venugopalan, Phys.\ Rev.\ D
{\bf 59}, 094002 (1999).
\end{thebibliography}
\end{document}